\documentclass[11pt]{article}

\usepackage[english]{babel}
\usepackage[letterpaper,top=2cm,bottom=2cm,left=3cm,right=3cm,marginparwidth=1.75cm]{geometry}

\usepackage[version=4]{mhchem} 
\usepackage{siunitx}
\usepackage{graphicx}% Include figure files
\usepackage{dcolumn}% Align table columns on decimal point
\usepackage{bm}% bold math
\usepackage{setspace}
\usepackage{amsmath}
\usepackage{graphicx}
\usepackage{fixmath}

\title{Depth-resolved Laue microdiffraction with coded-apertures}
\author{Do\u ga G\" ursoy\footnote{E-mail: dgursoy@anl.gov}, Dina Sheyfer, Michael Wojcik, Wenjun Liu and Jonathan Z. Tischler \\
X-ray Science Division, Argonne National Laboratory, \\
9700 S Cass Ave, Lemont, IL 60439, USA}

\begin{document}
\doublespacing
\maketitle

\begin{abstract}
We introduce a rapid data acquisition and reconstruction method to image the crystalline structure of materials and associated strain and orientations at micrometer resolution using Laue diffraction. Our method relies on scanning a coded-aperture across the diffracted x-ray beams from a broadband illumination, and a reconstruction algorithm to resolve Laue microdiffraction patterns as a function of depth along the incident illumination path. This method provides a rapid access to full diffraction information at sub-micrometer volume elements in bulk materials. Here we present the theory as well as the experimental validation of this imaging approach. 
\end{abstract}

\section{Introduction}

In crystalline materials such as metals, ceramics, alloys or polymers, the microstructural defects or deformations of the crystal lattice play an important role in defining the physical properties at macroscopic scales \cite{Wolf:92}. Transmission or scanning electron microscopy  \cite{Fundenberger:03,Wilkinson:97} can image the structure with atomic and nanometer resolution, however, usually crystalline materials are solids with long-range order and they highly absorb electrons, therefore only surfaces or thin materials on the order of a few micrometers can be studied with electrons. Retrieving the three-dimensional (3D) structure of bulk crystalline materials non-destructively has been the motivation behind several methods developed with x-rays, thanks to their unique properties, including high penetration depth and relative lack of multiple scattering \cite{Klug:74}. Especially with the advancement of undulator sources in the past decade, Laue microdiffraction using a sub-micrometer focused beam from a synchrotron storage ring has been established as a powerful technique to image crystalline strain, defects, deformations with sub-micrometer resolution in 2D and 3D \cite{ice20003d, kunz2009dedicated, tamura2002submicron, ulrich2011new, hofmann2009probing, maabeta2006defect, park2007local}.

\begin{figure}
\label{fig1}
\centering
\includegraphics{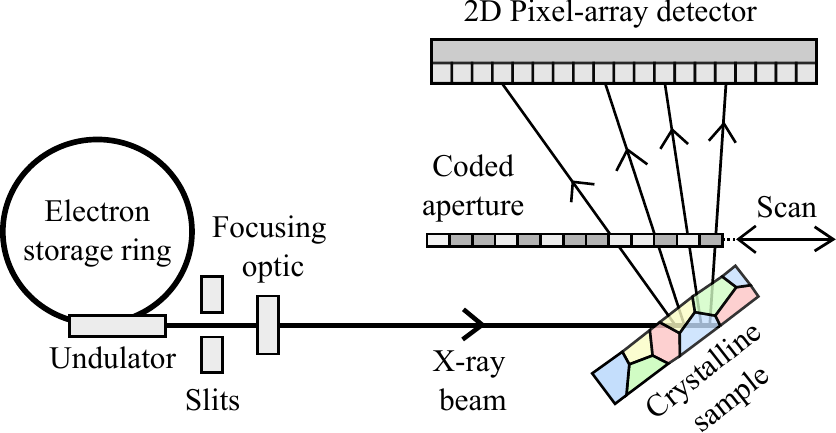}
\caption{Schematic representation of a coded Laue microdiffraction data collection. A crystalline sample is probed by an x-ray beam produced by synchrotron radiation available from wiggler and undulator sources passing through a micro-focusing optic, and the diffraction by crystallites along the beam path is recorded using a pixel-array detector mounted in a \SI{90}{\degree} reflection geometry. Scanning a coded-aperture allows encoding the diffracted beams in parallel.}
\end{figure}

In a typical Laue microdiffraction experiment, a focused polychromatic beam of x-rays is diffracted by individual crystallites along the illuminated path to the specific directions satisfying Bragg's law (Fig.~\ref{fig1}) \cite{chen2016quantitative}. A pixel array detector is used to record the diffracted signals from which one can infer information about the local structure of the crystal lattice and associated strain and orientations. A two-dimensional (2D) map of crystalline microstructure such as orientation, strain or deformation can be obtained by scanning a focused x-ray beam across the sample while Laue patterns are taken at each scanning point. Imaging the structure in 3D requires resolving the diffracted intensity along the axis of illumination. Depth resolution can be achieved through scanning an absorbing structure such as a wire or other fairly discontinuous edge as a differential-aperture to locate the diffracted signals one by one, and using ray tracing analysis to register those signals to their corresponding depths in the beam path \cite{Larson:02}. The method is analogous to a translating pinhole camera, and can provide depth-resolved diffraction information from sub-micrometer volume elements in material samples. 

One major limitation for performing experiments with the current 3D Laue microdiffraction instruments at synchrotrons is the long data acquisition time. Especially for \textit{in situ} or \textit{operando} studies in which the data is acquired multiple times while changing a control parameter such as temperature, pressure or other control parameters, the experiments can take many hours to days for achieving micrometer resolution of even \SI{50}{\micro\meter} sized samples. This is primarily because of the low signal-to-noise ratio (SNR) from a differential aperture coupled with the large number of raw images that contain no signal for a pixel but still contribute noise. Although multiple absorbing structures can be used to accelerate the acquisition process \cite{chung2007multiple}, accelerations beyond a factor of two or three is still unlikely due to the minimum spacing requirement of these structures in the aperture so as not to interfere with each other. Here, we propose using coded-aperture imaging, a well established technique in x-ray astronomy \cite{dicke1968scatter}, as an alternative to significantly reduce the scan range, and in turn the experiment time, while still obtaining a reconstruction of comparable quality. Our approach to measurement formation is to replace the scanning discontinuous structure with a coded-aperture, and to use an optimization based reconstruction algorithm for decoding the depth-resolved signals from the encoded measured images. Because scanning a coded-aperture modulates all the diffracted signals in parallel, a small scan range is sufficient for a successful depth recovery of signals.

\section{Methods}

\subsection{Measurement model}

To set the stage for the coded-aperture imaging technique, we first introduce the measurement process. Assume we have a one-dimensional (1D) coded-aperture that is described by the function $c(x)$. For a coded-aperture position $t$, we can express the intensity measurement $d_t$ of a scattered beam into a detector pixel as follows,
\begin{equation}
\label{eq0}
    d_t = \int_{\Omega} s(x) c(x - t) dx
\end{equation}
where $s(x)$ is the signal footprint along the beam scattered into the pixel in the coded-aperture plane, and $\Omega$ is the interval in which the signal is non-zero. Because the diffracted beams pass through different parts of the coded-aperture, translating the coded-aperture along the $x$-axis yields a unique encoding in measurements that we will leverage to resolve their position in $x$. 

To simulate the measurements, we discretisize the functions $c(x)$ and $s(x)$ with piecewise-constant functions on an equidistant grid. For a binary coded-aperture that uses two bits for the encoding, we can discretisize $c(x)$ and describe it by the vector $[a_0, a_1, \hdots, a_L]^T$ with each coefficient is optical transmissivity. For example, a $0$ means that an incident beam on the coded-aperture can pass through without being absorbed and a $1$ means that the signal is fully absorbed. In a realistic setting, the values are often between $0$ and $1$ and are dependent on the thickness and material type of the coded-aperture, as well as the energy and incidence angle of the incoming beam. In the discrete setting, we can express the measurement data $\mathbold{d} = [d_1, d_2, \hdots, d_M]^T$ in a single detector pixel as a result of translating the coded-aperture $M$ times (e.g. in equidistant steps) in terms of a matrix product,
\begin{equation}
\label{eq1}
\begin{bmatrix}
    d_{1} \\
    d_{2} \\
    \vdots \\
    d_{M}
\end{bmatrix}
=
\begin{bmatrix}
    a_{p} & a_{p+1} & \dots  & a_{p+N-1} \\
    a_{p+1} & a_{p+2} & \dots  & a_{p+N} \\
    \vdots & \vdots & \ddots & \vdots \\
    a_{p+M} & a_{p+M+1} & \dots  & a_{p+M+N-2}
\end{bmatrix}
\begin{bmatrix}
    s_{1} \\
    s_{2} \\
    \vdots \\
    s_{N}
\end{bmatrix},
\end{equation}
or simply in compact matrix notation as,
\begin{equation}
\label{eq2}
    \mathbold{d} = \mathbold{A}_p \mathbold{s}
\end{equation}
where $\mathbold{s} = [s_1, s_2, \hdots, s_N]^T$ is the signal footprint along the beam scattered into the pixel, $\mathbold{A}_p$ is the coding matrix, and $p$ is the offset scan position that represents the section of the coded-aperture (e.g. $[a_p, a_{p+1}, \hdots, a_{p+N-1}]^T$ at the first scan point) that modulates $\mathbold{s}$. $N$ is an open parameter that we can choose freely. A small $N$ is computationally desirable but it must be larger than the maximum signal size. Setting it to the largest probed depth would be the safest choice when we have no assumption about the nature of signals.

While a coding matrix with binary coefficients is ideal for optimally encoding the signals, it is often impractical due to the high penetration power of hard x-rays. However, a high atomic number material can still be used for the coding. For example, an x-ray energy range of 10-20~keV would require \SI{33}{\micro\meter} of gold (\ce{Au}) for 99\% absorption, while \SI{20}{\micro\meter} of \ce{Au} would absorb 95\% of x-rays. Thinner or different materials such as tungsten or platinum can as well be used to optimize the trade off between signal modulation contrast and noise in the encoded measurements. 

\subsection{Image reconstruction model}

The encoded signal collected in each detector pixel as a result of scanning the coded-aperture corresponds to a unique sub-string in the code sequence. To decode the signals projected by the coded-aperture, we need to find both the footprint of the diffracted signal and its position. We can achieve this by solving the following minimization problem of the form,
\begin{equation}
\label{eq3}
    \min_{p, \mathbold{s}} \left\| \mathbold{A}_p \mathbold{s} - \mathbold{d} \right\|_2^2,
\end{equation}
jointly for $p$ and $\mathbold{s}$. While there are many choices for solving Eq.~\ref{eq3}, we use a sequential optimization approach in which we update $p$ by fixing $\mathbold{s}$, followed by an update of $\mathbold{s}$ by fixing $p$. An alternating approaches by iteratively updating $p$ and $\mathbold{s}$ can as well be used for improved performance. To maximize the sensitivity to signal modulation contrast, we work with bias corrected and normalized signals, for example, $\mathbold{d} = (\mathbold{d}_{raw}-d_{min})/(d_{max}-d_{min})$ using the maximum $d_{max}$ and the minimum $d_{min}$ signals in the raw scan data $\mathbold{d}_{raw}$. When the incident intensity is unstable, we can consider using an additional measurement setup to monitor the incident beam and weight the measurements proportionally as part of the normalization procedure. 

Because the objective function in Eq.~3 is non-convex for a fixed $\mathbold{s}$, we solve the first problem of finding $p$ by an exhaustive search in which we check every signal position in a feasible range. In other words, we use a reasonably selected $\mathbold{s}$ and compute the objective function for all possible candidates of $p$ from the beginning of the coded-aperture to the end and pick the one that provides the minimum value. The algorithm can tolerate an inaccurate selection of $\mathbold{s}$ unless the sample has a very complicated crystalline structure. For example, a boxcar function about the average expected grain size can be used for $\mathbold{s}$. For the cases, when more complex signals are involved, then we may employ an alternating optimization strategy that can refine the signal and its position iteratively. Also when we calibrate the position of the coded-aperture before the experiment, we may have an expectation about the search range for the position, especially when the coded-aperture is close to the sample. This can help reduce the computational time to resolve the positions, but has no effect on the accuracy.

The following problem of finding $\mathbold{s}$ for a given $p$ is essentially a non-blind deconvolution problem in which we can solve directly with matrix inversion or possibly faster with an iterative solver. When $M$ is larger than $N$ (more data than the unknowns), the equation set is overdetermined, so an exact solution doesn't exist, but we can obtain an approximate solution with the least-squares method. The least-squares also offers attractive properties such as robustness to noise and wide availability of implementations. A conservative choice in the absence prior information about the sample is to collect as many measurements as the number of unknowns to describe the signal we want to recover and to ensure robustness by collecting high SNR measurements. However, when $M$ is smaller than $N$ or when we need to combat with noise, we usually incorporate additional constraints such as non-negativity or smoothness to limit the solution in a feasible set by effectively constraining the solution space.

When we recover the signal and its position on the coded-aperture plane, we can trace the signal from the detector pixel location back to its originating depth along the incident beam path. When we complete the ray-tracing process for all the pixels, we obtain the depth-resolved signals along the beam. The process is computationally cheap and requires extending the line described by the detector pixel position and the resolved position in the coded-aperture plane to the incident beam path. Because the diffracted beam can be detected at any point within the pixel, the certainty of the ray tracing direction is fundamentally limited by the detector pixel size. The effect leads to blurring and reduced depth resolution proportional to the distance from the coded-aperture to the incident beam path. In this paper, we assume that the detector pixel location is represented by its mid-point, but one can quantify the associated uncertainty in depth-resolved signals by modeling the blurring effect through a convolution, or more accurately, through segmenting the detector pixel with a grid structure and performing ray-tracing from multiple regions in the detector grid.

\subsection{Design and fabrication of the coded-aperture} 

\begin{figure}
\centering
\includegraphics{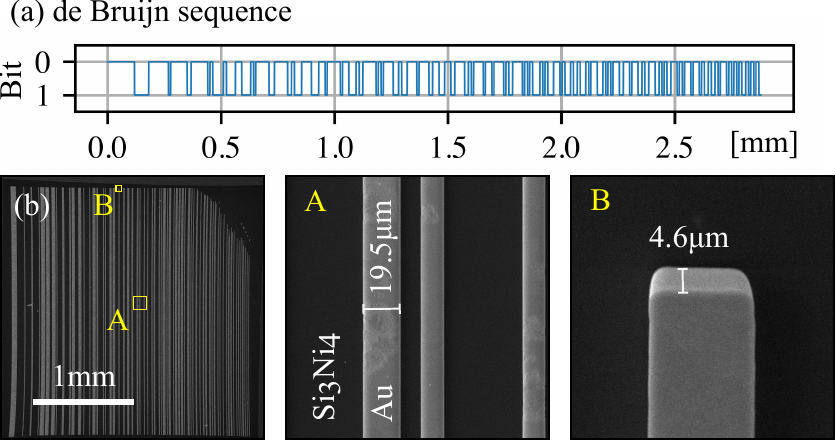}
\caption{(a) The code sequence used for the coded-aperture. (b) Scanning electron microscope image of the fabricated coded-aperture together with zoomed-in regions to highlight the individual bits, materials and thicknesses. }
\label{fig2}
\end{figure}

While there are many choices of codes for the coded-aperture, we specifically select de Bruijn sequences \cite{DeBrujin:46} as the basis for our aperture. De Bruijn sequences refer to a class of nonlinear cyclic sequences, in which every possible string of a particular length occurs exactly once as a substring. This key feature removes the decoding ambiguity, and allows us to uniquely resolve the Laue patterns originated from every depth along the beam path. For generating the coding sequence, we use a greedy algorithm \cite{gabric2018framework} that yields the shown binary de Bruijn sequence of order 8 and length 256 in Fig.~\ref{fig2}. In other words, every 8-bit sequence of our 256-bit long aperture is unique. The widths of the 1\,s and 0\,s in the code are chosen as \SI{7.5}{\micro\meter} and \SI{15}{\micro\meter}. This provides a minimum scan distance of about $8\times\SI{15}{\micro\meter}=\SI{120}{\micro\meter}$ for a unique encoding of all the signals. The coded-aperture was fabricated at the Center for Nanoscale Materials at Argonne National Laboratory. Direct-write lithography was used to generate the 2D barcode-like pattern from the sequence. A thin silicon nitride (\ce{Si3N4}) membrane (about $3 \times \SI{3}{\micro\meter}$) with \ce{Au} layer, needed for electroplating, was used as the aperture substrate. Several micrometers thick, positive resist was spun on the substrate and the pattern was written directly onto the membrane using lithography. The resulting mold was filled with \ce{Au} via electroplating, afterwards the remaining resist was removed using a solvent. A scanning electron microscope image of the fabricated coded-aperture is shown in Fig.~\ref{fig2} together with a zoomed in region to highlight several of the bits. Note that because of an imperfect fabrication process, all the \ce{Au} (1-bit) bars of our coded-aperture are about \SI{4.5}{\micro\meter} wider than the intended width of 7.5. The thickness of the fabricated coded-aperture is measured to be around \SI{4.6}{\micro\meter}.

\subsection{Experimental setup for data collection}

To validate our coded-aperture imaging method, we performed an experiment at the Laue microdiffraction instrument at the 34-ID-E beamline of Advanced Photon Source (APS) at Argonne National Laboratory. We used a polycrystalline nickel (\ce{Ni}) foil with a thickness of about \SI{75}{\micro\meter} as the imaging target. A polychromatic x-ray beam with energies in the range of 7-\SI{30}{\kilo\electronvolt} was focused using a pair of non-dispersive Kirkpatrick-Baez mirrors to a spot of approximately $300\times \SI{300}{\nano\meter}^2$. The sample was mounted at a \SI{45}{\degree} angle towards the incident x-ray beam, which gives sample depth range along the beam direction of about \SI{100}{\micro\meter}. A pixel-array Perkin-Elmer detector ($409.6\times\SI{409.6}{\milli\meter}^2$, $2048\times2048$ pixels, and 16-bit dynamic range) mounted in a \SI{90}{\degree} reflection geometry \SI{510.9}{\milli\meter} above the sample was used for acquiring the Laue patterns. The detector and coded-aperture scanning geometry with respect to the incident beam was calibrated using a strain-free thin silicon single crystal. The coded-aperture was placed about \SI{1}{\milli\meter} above the sample and was scanned with a \SI{1}{\micro\meter} step size along the beam direction. We use a step-scan and choose the exposure time per scan point \SI{0.25}{\second}. To obtain a depth-resolved 2D image of the sample, we repeated the scanning procedure at 31 sample positions for a total distance of \SI{155}{\micro\meter} at \SI{5}{\micro\meter} intervals.

\section{Results}

\begin{figure}
\centering
\includegraphics{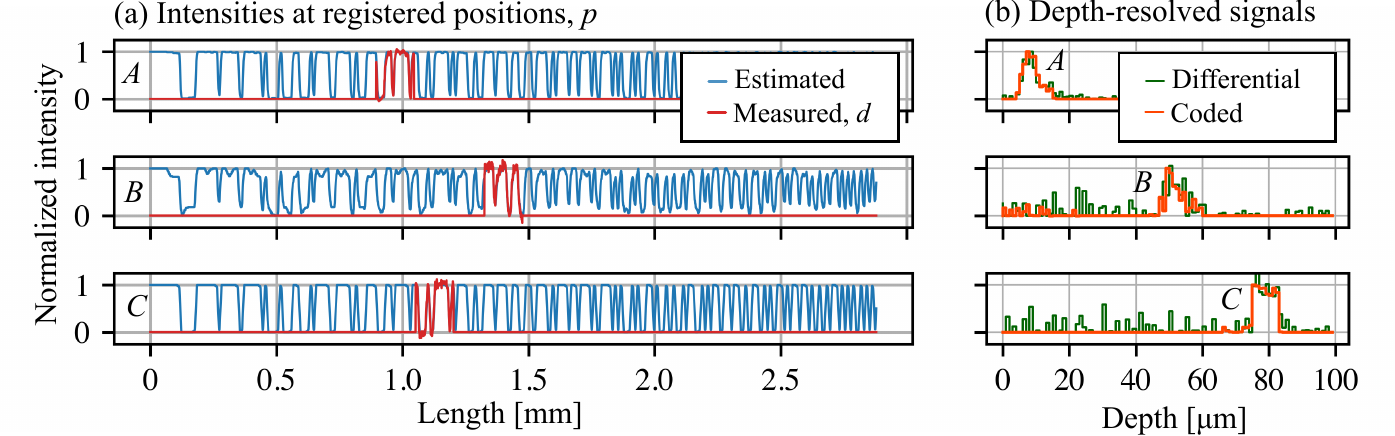}
\caption{Reconstruction results for three selected pixels in a measured Laue diffraction pattern. (a) Measured intensities in the shown pixels are aligned with their resolved position on the coded-aperture. The estimated signals are obtained by convolving the resolved signals with the coded-aperture. (b) Depth-resolved signals for each pixel using the coded-aperture and the differential-aperture imaging methods for the same \ce{Ni} sample location.}
\label{fig3}
\end{figure}

The depth-resolved Laue patterns were obtained by solving Eq.~\ref{eq3} for each pixel in the detector. We used a sequential approach by first recovering $p$ with an exhaustive search, and then recovering $\mathbold{s}$ with a non-negative least-squares solver \cite{lawson1995solving}. We show the results of this process of recovering $p$ and $\mathbold{s}$ in Fig.~\ref{fig3} for three selected detector pixels at the position. We plot the measured and estimated signals based on the recovered $p$ and $\mathbold{s}$ from which we observe that they show a good agreement. Note that the estimated signals are obtained by convolving the footprints $\mathbold{s}$ with the coded-aperture, and the measurement signals are aligned to their recovered positions $p$ on the coded-aperture. As a final step, with recovered $\mathbold{s}$ and $p$, we reconstruct the depth-resolved signals along the beam path using ray-tracing analysis as similar to the differential-aperture imaging method \cite{Larson:02}. We validate our reconstructions with the conventional differential-aperture method for the same \ce{Ni} sample location. For the validation scan, we use a \SI{100}{\micro\meter} platinum wire scanned with a \SI{1}{\micro\meter} step size for a total distance of \SI{600}{\micro\meter} across the sample with the same exposure time per scan point as the coded-aperture scanning. While both the differential-aperture and the coded-aperture method could reconstruct the signals at all depths, the differential-aperture reconstructions yield noisier signals even though it involves 5 times more scan points. In contrast, the coded-aperture is found to be more robust to noise induced in the depth-resolved signals during the reconstruction process. This reduction in noise is probably due to the increase in the number of edges that pass through each ray which increases the signal, and the fewer number of raw images most of which contain no signal for a particular pixel but still contribute noise. 

\begin{figure}
\centering
\includegraphics[width=\textwidth]{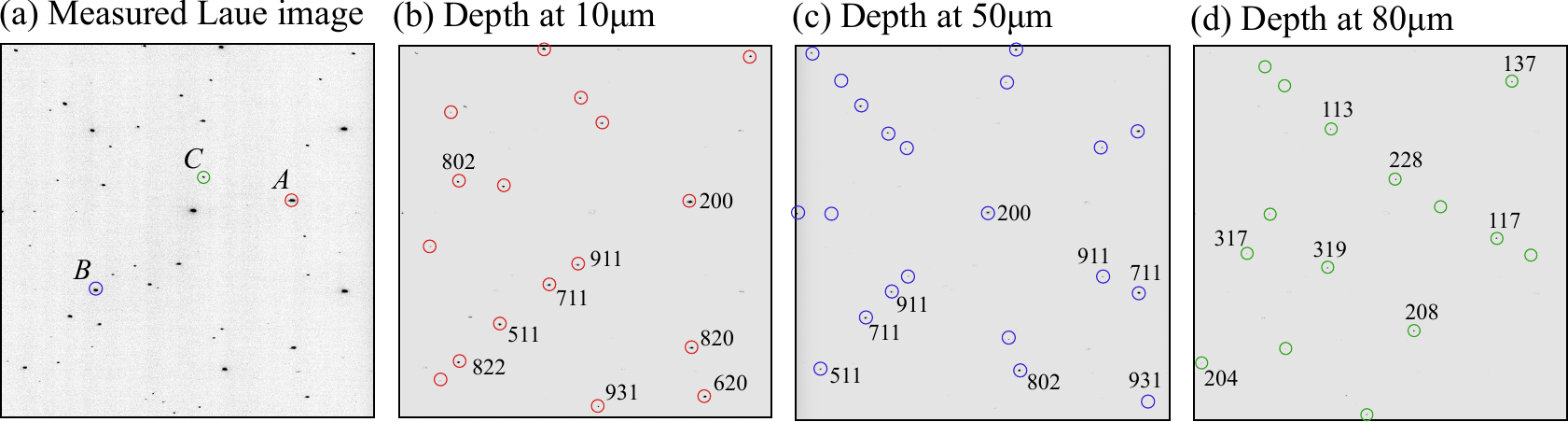}
\caption{(a) Measured Laue diffraction image that contains all the Bragg spots along the x-ray beam. \textit{A}, \textit{B} and \textit{C} show the locations of the pixels that are used for the analysis presented in Fig.~\ref{fig4}. (b-d) Decoded Laue diffraction patterns at selected depths along the beam. Resolved \textit{hkl} indices of the lattice are highlighted with circles.}
\label{fig4}
\end{figure}

By resolving the signals for all the detector pixels, the depth-resolved Laue diffraction patterns can be obtained, each corresponding to scattering from one micrometer volume element in the sample. We used LaueGo \cite{liu2014x}, an in-house developed software at the APS, for the analysis of those patterns. A number of indexed depth-resolved Laue patterns are presented in Fig.~\ref{fig4}. The corresponding \textit{hkl} indices of the crystal lattice are labeled in the images. We also label with circles the locations of the pixels \textit{A}, \textit{B} and \textit{C} that are used for the analysis in presented in Fig.~\ref{fig3}.

\begin{figure}
\centering
\includegraphics{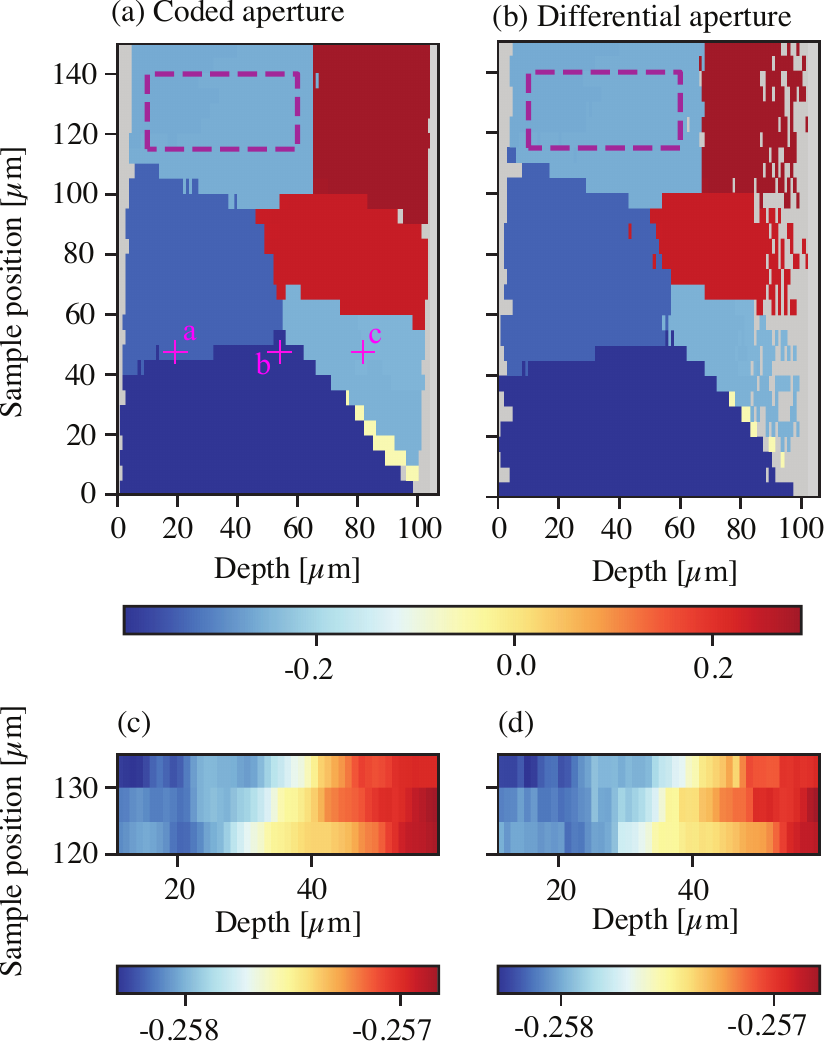} 
\caption{Depth-resolved reconstruction results of crystal lattice orientations with the coded-aperture (a) and the differential-aperture (b) imaging methods on the same \ce{Ni} sample. Color scale represents the rotation of crystal lattice in the units proportional to degrees. The plus markers denote the location of the Laue patterns shown in Fig~\ref{fig4}b, c and d, respectively. The orientation maps of a selected rectangular region in a single grain are shown in (c) and (d) for the coded- and differential-aperture methods, respectively.}
\label{fig5}
\end{figure}

The indexing results can be used to image different diffraction information about the sample. For example, in Fig.~\ref{fig5} we show the reconstruction of the crystal lattice orientations. The positions of the Laue patterns in Fig.~\ref{fig4} are labeled with plus markers. We also validate the coded-aperture reconstructions with the conventional differential-aperture reconstructions. The differential-aperture method yielded noisier reconstructions at depths beyond \SI{85}{\micro\meter}. In contrast, the coded-aperture imaging could recover the full extent of the sample, which is measured to be around \SI{100}{\micro\meter}. 

\begin{figure}
\centering
\includegraphics{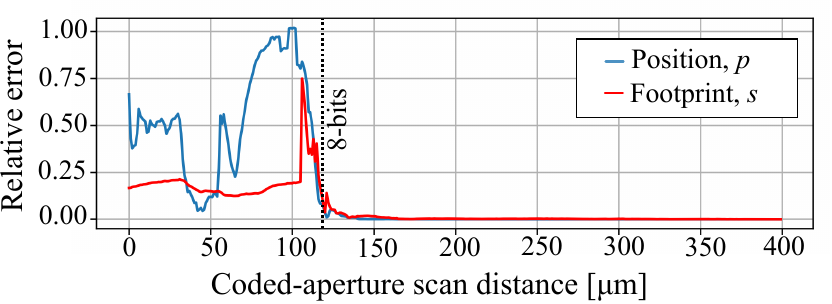}
\caption{Reconstruction error for different scan distances of the coded-aperture. While there might be ambiguities for smaller distances than the required 8 bits, the error is marginal when the aperture is scanned beyond 8 bits ($8\times15=120\si{\micro\meter}$) of distance.}
\label{fig6}
\end{figure}

Finally to evaluate the requirement for minimum scanning distance, we calculated the error $e[M]$ for different scan sizes $M$ relative to the maximum scan distance ($\text{max}=400~\si{\micro\meter}$). This can be expressed as follows,
\begin{equation}
    e[M] = \sum_i\frac{1}{d_i}\left(p_i[M] - p_i[\text{max}]\right)^2,
\end{equation}
where $p_i[M]$ is the reconstructed value of $i^{th}$ pixel by using a total distance of \SI{400}{\micro\meter} taken in \SI{1}{\micro\meter} steps, and $d_i$ is the intensity of that pixel. As can be seen in Fig~\ref{fig6}, we observe that the error is significant for small distances because of the ambiguity in the decoding process in which two or more substrings can provide the same signal. While it is probable that we may achieve a small error for smaller distances (e.g. at about \SI{45}{\micro\meter}), scanning a distance of 8 bits or more ensures a successful depth recovery. This observation also aligns well with the basic working principle of the method.  

\section{Discussions}

While the main advantage of using coded-apertures is the rapid data collection, it may as well be more robust to measurement noise and provide resolution at greater depths compared to the traditional differential-aperture approach. This is because with a differential-aperture, only two consecutive intensity recordings in a scan contribute to the reconstructed signal at a particular depth and the remaining intensities at other scan positions contribute to noise for that particular depth. Therefore, increasing the scan range has no effect on the quality of the signal recovery. A coded-aperture, on the other hand, can modulate the signal throughout the full course of the scan. Thus, all the recorded intensities in a pixel contribute to the signal resolution and increasing the scan range enable both improving the accuracy of the signal recovery and resolving signals at greater depths.

One key limitation of the proposed technique is that the coded-aperture position registration recovery process fails for SNR below a certain level. This may happen when the material of the coded-aperture is less absorbing or when the diffracted signal is weak due to a small crystallite or a dim source. In addition, absorption of both the incident and the diffracted beams hinder the accuracy of signal recovery at increasing depths, and beyond a certain depth we fail to recover the signals completely because all the diffracted signals along the beam path are absorbed by the sample. For imaging thicker or more absorbing samples, we think that this limitation can be relieved when we use realistic measurement noise models for solving Eq.~\ref{eq2} such as based on calculating the maximum likelihood or maximum a posteriori estimates \cite{Gursoy:15}. Use of single-photon counting detectors such as EIGER [21] can further relax this bottleneck associated with the low levels of measurement noise, however, the parallax error \cite{wernick2004emission} resulting from the lack of depth of interaction or incorrect registry of photons to detector pixels at oblique angles of incidence can lead to reduced resolution unless we account for these sources of error in the measurement model. 

Large signal footprints poses another challenge to consider, especially when there are large crystallites in the sample. Ideally, resolving a diffracted signal originating from a point source is the most straightforward, but usually, the diffraction occurs from a finite length along the beam path. As a result, the modulation in the measured signal becomes more convoluted with increasing crystallite size. A mild representation of this issue is readily observable in Fig.~\ref{fig3} in which the maximum and the minimum of the intensities don't reach the 1 and 0 value levels. In this experiment, the signal footprint sizes were about 10-\SI{30}{\micro\meter} in size and some of them were larger than the code bits, which were 7.5 and \SI{15}{\micro\meter}. While the substrings in the convoluted pattern still remain unique, the modulation contrast may be reduced for parts of the substring and the overall code structure can be distorted. Recovering larger crystallites with the same coded-aperture may probably require improved SNR in measurements or another optimizer that is more effective in this setting. For instance, an optimizer trained using supervised learning or a coded-aperture with larger spacing provide an excellent opportunity for future development.

The basic working principle of the proposed approach can be extended to 2D coded-apertures for further reductions in scanning. This requires using a coded-aperture with a checkerboard-like 2D pattern (e.g., a de Bruijn torus) in which each square-bit in the aperture is either an absorbing or a non-absorbing structure. In this setting, we scan the aperture not only in a single direction but in a 2D plane to track the measured signal back along a line directed towards the sample. In return, a vertically planar illumination instead of a point source can be used which excites more crystallites simultaneously. Although the fluence on the sample is not changed by converting a point source into a planar source, this approach trades the signal quality for faster acquisition. This 2D acquisition strategy may offer benefits when we have a bright radiation source, such as those from fourth-generation synchrotron radiation, or are limited by the scanning speed, for example due to mechanical reasons.

To fully exploit the high brilliance of synchrotron radiation sources, one needs to change from a step-and-shoot data acquisition to on-the-fly data acquisition by scanning of the aperture continuously. With this approach we can eliminate the time (and photons) wasted in waiting for the mechanical settling of our scanner system before recording each image, but a continuous scanning will induce motion blur in measurements. A reasonable approach to take is to operate the detector at a frame rate in which we still have acceptable SNR in measurements, and set the scan speed such that the blurring is negligible. For experimental conditions when faster scanning is desired, one can encode the signals in time. The working principle of motion correction with time-coded measurements is conceptually equivalent to coded-aperture imaging, except that the signals are modulated temporally instead of spatially \cite{Ching:19}.

Although we use a binary coded-aperture with a fixed thickness, a non-binary coded-aperture with varying thicknesses can be used to simultaneously encode the energy information of the diffracted beam. This can be achieved by using a larger alphabet size for generating the de Bruijn sequence, and fabricating a coded-aperture in which each letter in the aperture can be described with a varying aperture thickness. For example, a de Bruijn sequence $\{0, 0, 1, 1, 2, 2, 3, 3, 0, 2, 0, 3, 1, 3, 2, 1\}$ can be constructed in which 0, 1, 2, and 3 mean 0 (no \ce{Au}), 5, 10, \SI{20}{\micro\meter} of \ce{Au} thickness. This would not only shorten the length for scanning further, but also provide the energy information about the diffracted beam, because the absorption properties of the aperture is varying as a function of beam energy. Therefore, with this method one can simultaneously estimate the energy profile of each pixel while scanning. 

In summary, we demonstrate the coded-aperture imaging method for Laue microdiffraction experiments at the synchrotrons. This pilot study provides encouraging results and suggest new opportunities for further development and optimization of coded-apertures, algorithms and scanning schemes.

\section*{Acknowledgements}
This research used resources of the Advanced Photon Source and the Center for Nanoscale Materials, U.S. Department of Energy (DOE) Office of Science User Facilities and is based on work supported by Laboratory Directed Research and Development (LDRD) funding from Argonne National Laboratory, provided by the Director, Office of Science, of the U.S. DOE under Contract No. DE-AC02-06CH11357.

\bibliographystyle{ieeetr}
\bibliography{main}

\end{document}